\input amssym.def
\input amssym
\documentstyle[10pt]{article}
\def\scs{\scriptsize}
\def\ol{\overline}
\def\tr{\mbox{Tr}}
\def\tp#1{#1^{\mbox{\scs{T}}}}
\def\dim{\mbox{dim}\;}

\def\com{\mbox{Comm}}
\def\a{\alpha}
\def\b{\beta}
\def\c{\gamma}
\def\d{\delta}

\def\s{\sigma}
\def\cB{{\cal B}}
\def\fa{\frak A}
\def\fb{\frak B}
\def\fc{\frak C}
\def\dg{\dagger}
\def\ds{\displaystyle}
\def\lra{\Leftrightarrow}
\def\dirsum#1{\mbox{{\normalsize ${\ds\bigoplus_{#1}}$}}}
\def\ix{{\rm ad,F,B}}
\newtheorem{defn}{Definition}
\newtheorem{prop}{Proposition}
\newtheorem{rem}{Remark}
\newtheorem{thm}{Theorem}
\newcommand{\be}{\begin{equation}}
\newcommand{\ee}{\end{equation}}
\newcommand{\bea}{\begin{eqnarray}}
\newcommand{\eea}{\end{eqnarray}}
\newcommand{\beax}{\begin{eqnarray*}}
\newcommand{\eeax}{\end{eqnarray*}}
\newcommand{\no}{\noindent}
\newcommand{\op}[1]{\mbox{\sf #1}}
\newcommand{\komm}[2]{\left[#1,#2\right]}
\newcommand{\akomm}[2]{\left\{#1,#2\right\}}
\newcommand{\dfrac}[2]{{\ds\frac{#1}{#2}}}
\newcommand{\bd}{\begin{defn}{\hspace{-.55em}\em{\bf {: }}}}
\newcommand{\ed}{\end{defn}}
\newcommand{\bp}{\begin{prop}{\hspace{-.55em}\em{\bf {: }}}}
\newcommand{\ep}{\end{prop}}
\newcommand{\br}{\begin{rem}{\hspace{-.55em}\em{\bf {: }}}}
\newcommand{\er}{\end{rem}}
\newcommand{\bt}{\begin{thm}{\hspace{-.55em}\em{\bf {: }}}}
\newcommand{\et}{\end{thm}}
\newcommand{\prbeg}{\no{\bf Proof: }}
\newcommand{\prend}{\hfill$\blacksquare$\\}

\begin{document}
\bibliographystyle{plain}
\begin{titlepage}
\renewcommand{\thefootnote}{\fnsymbol{footnote}}
\large
\hfill\begin{tabular}{l}HEPHY-PUB 653/96\\ UWThPh-1996-49\\ hep-th/9702137\\
December 1996
\end{tabular}\\[3cm]
\begin{center}
{\Large\bf CLIFFORD ALGEBRAS IN}\\[.5ex]
{\Large\bf FINITE QUANTUM FIELD THEORIES}\\[1cm]
{\Large\bf I.~Irreducible Yukawa Finiteness Condition}\\
\vspace{1.7cm}
{\Large\bf Wolfgang LUCHA\footnote[1]{\normalsize\ E-mail:
v2032dac@awiuni11.edvz.univie.ac.at}}\\[.5cm]
Institut f\"ur Hochenergiephysik,\\
\"Osterreichische Akademie der Wissenschaften,\\
Nikolsdorfergasse 18, A-1050 Wien, Austria\\[1.7cm]
{\Large\bf Michael MOSER\footnote[2]{\normalsize\ E-mail:
mmoser@galileo.thp.univie.ac.at}}\\[.5cm]
Institut f\"ur Theoretische Physik,\\
Universit\"at Wien,\\
Boltzmanngasse 5, A-1090 Wien, Austria\\[2cm]
{\bf Abstract}
\end{center}
\normalsize

Finite quantum field theories may be constructed from the most general
renormalizable quantum field theory by forbidding, order by order in the
perturbative loop expansion, all ultraviolet-divergent renormalizations of
the physical parameters of the theory. The relevant finiteness conditions
resulting from this requirement relate all dimensionless couplings in the
theory. At first sight, Yukawa couplings which are equivalent to the
generators of some Clifford algebra with identity element represent a very
promising type of solutions of the condition for one-loop finiteness of the
Yukawa couplings. However, under few reasonable and simplifying assumptions
about their particular structure, these Clifford-like Yukawa couplings prove
to be in conflict with the requirements of one- and two-loop finiteness of
the gauge coupling and of the absence of gauge anomalies, at least for all
simple gauge groups up to and including rank 8.

\vspace*{1ex}

\normalsize
\renewcommand{\thefootnote}{\arabic{footnote}}
\end{titlepage}

\section{Introduction}

Renormalizable quantum field theories appear to be the appropriate framework
for the comprehensive understanding of nature at a rather fundamental level.
In particular, the so-called ``standard model'' of elementary particle
physics, a spontaneously broken non-Abelian gauge theory based on the gauge
group ${\rm SU}(3)\times {\rm SU}(2)\times {\rm U}(1)$, describes extremely
successfully the strong and electroweak interactions. At present, this
standard model is jeopardized only by the still unsettled question of the
existence of the Higgs boson, required by the mechanism for spontaneous
breakdown of a (local) gauge symmetry. Nevertheless, renormalizable quantum
field theories exhibit, in general, the not very appealing feature that, in
their loopwise perturbative evaluation, there still appear ultraviolet
divergences, even though these can be handled by application of the
renormalization programme. Therefore, it is, beyond doubt, legitimate to
wonder whether among all the renormalizable quantum field theories there are
theories which are {\em finite\/}, in the sense that they do not evolve
ultraviolet divergences (up to some loop order).

Supersymmetry, by reducing the number of uncorrelated ultraviolet
divergences in quantum field theories, represents the first example of a
global symmetry which allows to construct finite quantum field theories:
\begin{itemize}
\item All one-loop finite $N=1$ supersymmetric theories are (at least)
two-loop finite \cite{N=1}, even if this $N=1$ supersymmetry is softly broken
(in a well-defined way) \cite{N=1-soft}. Under certain circumstances, $N=1$
supersymmetric theories may be finite to all orders of their perturbative
expansion \cite{piguet}.
\item All $N=2$ supersymmetric theories satisfying merely one single
``finiteness condition'' are finite to all orders of the perturbative
expansion \cite{N=2}, even if one or both supersymmetries are softly broken
(in a well-defined way) \cite{N=2-soft}; these theories have been classified
under various aspects \cite{N=2-class}.
\item In the case of the $N=4$ supersymmetric Yang--Mills theory, that ``$N=2$
finiteness condition'' is trivially fulfilled by the particle content of this
theory enforced by $N=4$ supersymmetry \cite{N=4}.
\end{itemize}

Clearly, the next logical step is to impose the requirement of finiteness to
arbitrary renormalizable quantum field theories in four space-time dimensions
\cite{lucha86-1,lucha86-2}; of particular interest here is the question
whether every finite theory must indeed be supersymmetric. The inspection of
general gauge theories shows immediately that finiteness of some quantum
field theory may only be achieved if the particle content of this theory
comprises {\em vector bosons, fermions, and scalar bosons\/}
\cite{lucha86-1,lucha86-2,boehm87,lucha87-1}. The complete set of finiteness
conditions for general quantum field theories has not yet been solved. Some
insights, however, may be gained by analysis of specific (classes of) models.
For instance, models being finite in dimensional regularization, at least up
to some loop order, may be shown to be plagued by quadratic divergences in
cut-off regularization \cite{qdiv1,qdiv2}.

A useful instrument in the search for non-supersymmetric finite theories is
the observation \cite{kranner90,kranner91} that, for all finite quantum field
theories, a certain group-theoretic quantity turns out to be bounded. In
fact, it has even been speculated \cite{kranner91} that all finite theories
might belong to a particular class of models characterized by the
circumstance that this group-theoretic quantity takes its maximal value.
Within this class---which encompasses all supersymmetric finite models
\cite{kranner91}---attempts to construct explicit non-supersymmetric finite
theories have been undertaken \cite{skarke94-2} and large sets of such
candidate models based on the gauge group SU($N$) have been excluded
\cite{grandits}.

In the course of analyzing this specific class of models, explicit solutions
of the one-loop finiteness condition for the Yukawa couplings which resemble
the generators of a Clifford algebra with identity element have been found
\cite{kranner91}. The present investigation scrutinizes the relevance of
these Clifford-like Yukawa solutions for the construction of new, i.e.,
non-supersymmetric, finite quantum field theories.

The outline of this paper is as follows: In Sec.~\ref{sec:fogqft}, we
formulate the conditions under which we regard an arbitrary quantum field
theory as finite (up to some loop order). For the investigation of the
high-energy behaviour of some quantum field theory, only the massless limit
of this theory, characterized by the vanishing of all dimensional parameters
in this theory, is relevant. Consequently, without loss of generality, we
confine ourselves to the discussion of theories involving only dimensionless
couplings. In the order of increasing complexity, the first genuine hurdle to
be taken is the condition for one-loop finiteness of the Yukawa couplings.
Finding corresponding solutions is greatly facilitated by adopting the
standard form of this relation, re-derived in Sec.~\ref{sec:yfc}. The
above-mentioned specific class of models is briefly reviewed in
Sec.~\ref{fin28a}. Stripping off irrelevant ballast, the one-loop Yukawa
finiteness condition is reduced, in Sec.~\ref{sec:reduc}, to its ``hard
core'' which, under the simplifying assumptions about the structure of the
Yukawa couplings specified in Sec.~\ref{sec:carfi}, is then carefully
investigated along the lines sketched in Sec.~\ref{sec:num}.
Section~\ref{sec:sco} summarizes our findings, the requirements for their
validity, and the way they may be obtained. Several more or less merely
technical details are banished to Appendices \ref{a1} through \ref{a12}.

\section{Finiteness of General Quantum Field Theories}\label{sec:fogqft}

The starting point of our considerations is the most general
\cite{llewellyn73} renormalizable quantum field theory (for particles up to
spin 1 $\hbar$) invariant with respect to gauge transformations forming some
compact simple Lie group $G$ with corresponding Lie algebra $\fa$. The
particle content of this theory consists of
\begin{itemize}
\item gauge vector bosons $A_\mu(x)=(A_\mu^a)(x)\in\fa$, transforming
according to the adjoint representation $R_{{\rm ad}}$: $\fa\rightarrow\fa$
of the gauge group $G$, of dimension $d_{\rm g}:=\dim\fa$;
\item two-component Weyl fermions $\psi(x)=(\psi^i)(x)\in V_{\rm F}$,
transforming according to a representation $R_{\rm F}$: $V_{\rm F}\rightarrow
V_{\rm F}$ of $G$, of dimension $d_{\rm F}:=\dim V_{\rm F}$; and
\item real scalar bosons $\phi(x)=(\phi^\a)(x)\in V_{\rm B}$, transforming
according to some real representation $R_{\rm B}$: $V_{\rm B}\rightarrow
V_{\rm B}$ of $G$, of dimension $d_{\rm B}:=\dim V_{\rm B}$.
\end{itemize}
Apart from terms involving dimensional parameters, like mass terms and cubic
self-interaction terms of scalar bosons, as well as gauge-fixing and ghost
terms, the Lagrangian defining this theory is given by
\bea
\cal L&=&-\frac{1}{4}\,F_{\mu\nu}^a\,F^{\mu\nu}_a
+i\,\ol{\psi}_i\,\ol{\s}^\mu\,(D_\mu)_{\rm F}\,\psi^i
+\frac{1}{2}\left[(D_\mu)_{\rm B}\phi^\a\right](D^\mu)_{\rm B}\phi_\a
\nonumber\\[1ex]
&&-\frac{1}{2}\,\phi^\a\,Y_{\a ij}\,\psi^i\,\psi^j
-\frac{1}{2}\,\phi_\a\,Y^{\dg\a ij}\,\ol{\psi}_i\,\ol{\psi}_j
-\frac{1}{4!}\,V_{\a\b\c\d}\,\phi^\a\,\phi^\b\,\phi^\c\,\phi^\d\ .
\label{fin1}
\eea
Here, we employ the following notation: The Hermitean generators $T_{\rm
R}^a$, ${\rm R} = \ix$, $a=1,2,\dots,d_{\rm g}$, of the Lie algebra ${\fa}$
in each of the three representations $R_{{\rm ad}}$, $R_{\rm F}$, and $R_{\rm
B}$ introduced above satisfy the commutation relations
\be
\komm{T_{\rm R}^a}{T_{\rm R}^b}=i\,{f^{ab}}_c\,T_{\rm R}^c\ ,\quad
{\rm R} = \ix\ ,
\label{fin4}
\ee
where ${f^{ab}}_c$, $a,b,c=1,2,\dots,d_{\rm g}$, denote the structure
constants characterizing the Lie algebra $\fa$. The gauge coupling constant
is denoted by $g$. The (gauge-covariant) field strength tensor $F_{\mu\nu}^a$
is given by
\be
F_{\mu\nu}^a=\partial_\mu A_\nu^a-\partial_\nu A_\mu^a
+g\,{f^a}_{bc}\,A_\mu^b\,A_\nu^c\ ,
\label{fin2}
\ee
The gauge-covariant derivatives $D_\mu$ acting on the representation spaces
$\fa$, $V_{\rm F}$, and $V_{\rm B}$, respectively, read
\be
(D_\mu)_{\rm R}:=\partial_\mu-i\,g\,T_{\rm R}^a\,A_\mu^a\ ,\quad
{\rm R} = \ix\ .
\label{fin3}
\ee
Finally, the four $2\times 2$ matrices $\ol{\s}^\mu$ embrace the $2\times 2$
unit matrix $\op{1}_2$ and the three Pauli matrices $\mbox{\boldmath$\s$}$
according to the definition $\ol{\s}^\mu=(\op{1}_2,-\mbox{\boldmath$\s$})$.

Quite obviously, the Yukawa couplings $Y_{\a ij}$ must be totally symmetric
in their fermionic indices $i$ and $j$, and the quartic scalar-boson
self-couplings $V_{\a\b\c\d}$ must be totally symmetric under an arbitrary
permutation of their indices.

In order to facilitate the formulation of the finiteness conditions below, we
would like to introduce some (group-theoretic) quantities. For an arbitrary
representation $R$ of $G$, we define, in terms of the generators $T_R^a$ of
${\fa}$ in this representation, the corresponding quadratic Casimir operator
$C_R$ by
\be
C_R:=\sum_{a=1}^{d_{\rm g}}T_R^a\,T_R^a
\label{fin5}
\ee
and the corresponding Dynkin index $S_R$ by
\be
S_R\,\delta^{ab}:=\tr\left(T_R^a\,T_R^b\right)\ .
\label{fin6}
\ee
In the adjoint representation $R_{{\rm ad}}$, the Casimir eigenvalue $c_{\rm
g}$ equals the Dynkin index $S_{\rm g}$, i.e., $c_{\rm g}=S_{\rm g}$.
Moreover, we shall take advantage of the abbreviations
\bea
Q_{\rm F}&:=&\sum_If_I\,S_I\,C_I\equiv\frac{1}{d_{\rm g}}\,\tr(C_{\rm F})^2\ ,
\nonumber\\[1ex]
Q_{\rm B}&:=&\sum_Ib_I\,S_I\,C_I\equiv\frac{1}{d_{\rm g}}\,\tr(C_{\rm B})^2\ ,
\label{fin7}
\eea
where the summation index $I$ runs over all inequivalent irreducible
representations $R_I$ of multiplicities $f_I$ and $b_I$ in $R_{\rm F}$ and
$R_{\rm B}$, respectively. Finally, it proves to be advantageous to
introduce the shorthand notation
\be
E(Y):=6\,g^2\,\tr_{\rm F}\left(C_{\rm F}\sum_{\b=1}^{d_{\rm
B}}Y^{\dg\b}\,Y_\b\right)\ ,
\label{eq:e(y)}
\ee
where by $\tr_{\rm F}$ we mean the partial trace over the fermionic indices
only.

With all the above preliminaries, we are now in the position to formulate the
finiteness conditions we are interested in. We adhere to the notion of
``finiteness'' for general renormalizable quantum field theories as advocated
and investigated first in Refs.~\cite{lucha86-1,lucha86-2}. Hence, any such
theory will be regarded as ``finite'' if it does not require divergent
renormalizations of its physical parameters, that is, masses and coupling
constants. This is equivalent to demanding finiteness of the resulting
$S$-matrix elements (not of the Green's functions) without divergent
renormalizations of the involved coupling constants. Consequently, our
finiteness conditions may be found by requiring the beta functions of these
physical parameters to vanish. Evidently, within a perturbative evaluation of
the quantum field theory under consideration, the vanishing of all beta
functions must take place order by order in the loop expansion. By
application of the standard renormalization procedure with the help of
dimensional regularization in the minimal-subtraction scheme, the relevant
finiteness conditions may be easily extracted \cite{cheng74,machacek}, see
also Refs.~\cite{lucha86-1,lucha86-2}; they read, for one-loop finiteness of
the gauge coupling constant $g$,
\be
22\,c_{\rm g}-4\,S_{\rm F}-S_{\rm B}=0\ ,
\label{fin8}
\ee
for two-loop finiteness of the gauge coupling constant $g$,
\be
E(Y)-12\,g^4\,d_{\rm g}\,[Q_{\rm F}+Q_{\rm B}+c_{\rm g}\,(S_{\rm
F}-2\,c_{\rm g})]=0\ ,
\label{fin9}
\ee
and, for one-loop finiteness of the Yukawa couplings $Y_{\a ij}$,
\bea
&&\sum_{\b=1}^{d_{\rm B}}\left\{4\,Y_\b\,Y^{\dg\a}\,Y_\b+Y_\a\,Y^{\dg\b}\,Y_\b
+Y_\b\,Y^{\dg\b}\,Y_\a+Y_\b\,\tr\left(Y^{\dg\a}\,Y_\b+Y^{\dg\b}\,Y_\a\right)
\right\}\nonumber\\[1ex]
&&-\;6\,g^2\left[Y_\a\,C_{\rm F}+\tp{\left(C_{\rm F}\right)}\,Y_\a\right]=0\ .
\label{fin10}
\eea
In the following, we call Eq.~(\ref{fin10}), our main concern, for short,
``Yukawa finiteness condition'' (YFC). It has been noticed at several
occasions in the literature \cite{kranner91,grandits} that the above
lowest-order finiteness conditions for gauge and Yukawa couplings, i.e.,
Eqs.~(\ref{fin8}), (\ref{fin9}), and (\ref{fin10}), constitute the central
part of the whole set of finiteness conditions, in the sense that the
inspection of the finiteness conditions for the quartic scalar-boson
self-couplings $V_{\a\b\c\d}$ or of higher order in the loop expansion makes
sense only after this central part has been solved.

\section{The Standard Form of the Yukawa Finiteness Condition}\label{sec:yfc}

Let $B_{\rm F}=\{e_i\}$ be some basis of the ``fermionic'' representation
space $V_{\rm F}$ and let $B_{\rm B}=\{f_\a\}$ be some basis of the
``bosonic'' representation space $V_{\rm B}$; in terms of these bases, we may
write
\beax
\psi&=&\sum_{i=1}^{d_{\rm F}}\psi^i\,e_i\ ,\\[1ex]
\phi&=&\sum_{\a=1}^{d_{\rm B}}\phi^\a\,f_\a\ .
\eeax
Then $Y_{\a ij}$ may be interpreted as the components of a {\em Yukawa
coupling tensor\/} $Y$ in the corresponding tensor basis $\{f_\a\otimes
e_i\otimes e_j\}$ of the product space $V_{\rm B}\times V_{\rm F}\times
V_{\rm F}$. Gauge invariance of the Lagrangian ${\cal L}$ requires the
invariance of $Y$ under the contragredient representation $R^{\rm c}$ of
$R=R_{\rm B}\otimes R_{\rm F}\otimes R_{\rm F}$:\footnote{\normalsize\ This
statement expresses, of course, nothing else but the (trivial) fact that the
Yukawa coupling strength for any fixed irreducible representations
$R_I,R_J\subset R_{\rm F}$ and $R_A\subset R_{\rm B}$ is not affected by
gauge transformations.}
\be
\left(R_{\rm B}^{\rm c}\otimes R_{\rm F}^{\rm c}\otimes R_{\rm F}^{\rm c}
\right)Y\equiv Y\ .
\label{fin12}
\ee

Let us now introduce a quantity $x=({x^{i\a}}_{j\b})$, which transforms like
an operator\footnote{\normalsize\ To be more precise, ${x^{i\a}}_{j\b}$ may
be interpreted as the components of the operator $x$ with respect to the
tensor basis $\{e_i\otimes f_\a\}$.} on the product space $V_{\rm F}\times
V_{\rm B}$, by defining
\be
2\,{x^{i\a}}_{j\b}={\left(Y^{\dg\a}\,Y_\b+Y^{\dg\b}\,Y_\a\right)^i}_j\ .
\label{fin14}
\ee
\bp
The operator $x$ is gauge invariant and diagonalizable on $V_{\rm F}\times
V_{\rm B}$.
\label{fin15}
\ep
\prbeg
Since $x$ is normal it is diagonalizable. The gauge invariance of $x$ is
shown in Appendix~\ref{a1}.
\prend

A system $\Sigma\ni M$: $V\rightarrow V$ of matrices is called {\em
reducible\/} if there exists an invariant subspace of $V$ under the action of
$\Sigma$, else $\Sigma$ is called {\em irreducible}. The commutant of such a
system $\Sigma$, defined by $\com(\Sigma):=\{\mbox{$N$:}\ V\rightarrow
V\mid\komm{M}{N}=0,\ \forall\ M\in\Sigma\}$, forms a matrix algebra
\cite{boerner55}. Now, suppose that $\Sigma$ is completely reducible, i.e.,
that $\Sigma$ is the direct sum of irreducible systems. In this case
$\com(\Sigma)$ is isomorphic to the direct sum of matrix rings
\cite{boerner55}. Let $M\in\Sigma$ and $N\in\com(\Sigma)$. We may write
\bea
M&=&\dirsum{i}\,\op{1}_{r_i}\times M_i\ ,\nonumber\\[1ex]
N&=&\dirsum{i}\,N_i\times\op{1}_{n_i}\ ,
\label{fin21}
\eea
where $i$ labels the inequivalent irreducible components $M_i$ of $M$, of
dimension $n_i$ and multiplicity $r_i$, respectively, $\op{1}_d$ represents
the $d$-dimensional unit matrix, and $N_i$ denotes an arbitrary $r_i\times
r_i$ matrix. In Ref.~\cite{kranner91} it was shown that the YFC is invariant
under an arbitrary $U(d_{\rm F})\otimes O(d_{\rm B})$ transformation. We may
take advantage of the $U(d_{\rm F})$ symmetry by choosing $B_{\rm F}$ such
that $R_{\rm F}$ becomes blockdiagonal in each irreducible representation
$R_{\rm F}^I$. The $O(d_{\rm B})$ symmetry may transform $R_{\rm B}$ into a
direct sum of real orthogonal blocks\footnote{\normalsize\ Every
non-orthogonal irreducible representation $R_{\rm B}^A\subset R_{\rm B}$ has
to find a mutually contragredient companion $(R_{\rm B}^A)^{\rm c}\subset
R_{\rm B}$ in order to be able to form a real orthogonal block: $R_{\rm
B}^\mu\simeq R_{\rm B}^A\oplus (R_{\rm B}^A)^{\rm c}$.} $R_{\rm B}^\mu=R_{\rm
B}^{\mu*}$:
\bea
R_{\rm F}&=&\dirsum{I}\,R_{\rm F}^I\ ,\nonumber\\[1ex]
R_{\rm B}&=&\dirsum{\mu}\,R_{\rm B}^\mu\ .
\label{fin21a}
\eea
For any operator acting on the product space $V_{\rm F}\times V_{\rm B}$, we
define, with respect to some corresponding tensor basis $\{e_i\otimes
f_\a\}$, partial traces $\tr_{\rm B}$ and $\tr_{\rm F}$ over bosonic and over
fermionic indices, respectively. For ${x^{i\a}}_{j\b}$, the contraction of
either the two bosonic or the two fermionic indices yields
\bea
{(y_{\rm F})^i}_j&:=&\sum_{\b=1}^{d_{\rm B}}{x^{i\b}}_{j\b}
=\sum_{\b=1}^{d_{\rm B}}\sum_{k=1}^{d_{\rm F}}Y^{\dg\b ik}\,Y_{\b kj}\ ,
\nonumber\\[1ex]
2\,{(y_{\rm B})^\a}_\b&:=&\sum_{i=1}^{d_{\rm F}}{x^{i\a}}_{i\b}
=\sum_{i,j=1}^{d_{\rm F}}\left(Y^{\dg\a ij}\,Y_{\b ji}+Y^{\dg\b ij}\,Y_{\a ji}
\right)
=\tr_{\rm F}\left(Y^{\dg\a}\,Y_{\b}+Y^{\dg\b}\,Y_{\a}\right)\ .
\label{fin22}
\eea
$y_{\rm F} = \tr_{\rm B} x$ and $y_{\rm B} = \tr_{\rm F} x$ transform as
invariant operators on $V_{\rm F}$ and $V_{\rm B}$, respectively. By
choosing, for every type of mutually equivalent blocks in $R_{\rm F}$ and
$R_{\rm B}$ a representative $R^I$ and $R^\mu$, respectively, we have, in the
notation (\ref{fin21}),
\bea
R_{\rm F}&=&\dirsum{I}\,\op{1}_{f_I}\times R^I\ ,\nonumber\\[1ex]
R_{\rm B}&=&\dirsum{\mu}\,\op{1}_{b_\mu}\times R^\mu\ ,
\label{fin23}
\eea
where the direct sums in $R_{\rm F}$ and $R_{\rm B}$ extend over all
inequivalent irreducible representations $R^I\subset R_{\rm F}$, of
dimensions $d_I$ and multiplicities $f_I$, as well as all inequivalent
orthogonal representations $R^\mu\subset R_{\rm B}$, of dimensions $d_\mu$
and multiplicities $b_\mu$, respectively. The invariance of $y_{\rm F}$ and
$y_{\rm B}$ under $R_{\rm F}$ and $R_{\rm B}$, respectively, implies $y_{\rm
F}\in\com(R_{\rm F})$ and $y_{\rm B}\in\com(R_{\rm B})$. According to
Eq.~(\ref{fin21}), these operators may be represented in the form
\bea
y_{\rm F}&=&\dirsum{I}\,W_I\times\op{1}_{d_I}\ ,\nonumber\\[1ex]
y_{\rm B}&=&\dirsum{\mu}\,Z_\mu\times\op{1}_{d_\mu}\ ,
\label{fin24}
\eea
for arbitrary systems of $f_I\times f_I$ matrices $W_I$ and $b_\mu\times
b_\mu$ matrices $Z_\mu$. The invariance of the operators $y_{\rm F}$ and
$y_{\rm B}$ under $R_{\rm F}$ and $R_{\rm B}$, respectively, guarantees the
vanishing of their commutators with the corresponding Casimir operators
$C_{\rm F}$ and $C_{\rm B}$:
\bea
\komm{y_{\rm F}}{C_{\rm F}}=0\ ,\nonumber\\[1ex]
\komm{y_{\rm B}}{C_{\rm B}}=0\ .
\label{fin25}
\eea
Now, for a finite or infinite system of diagonalizable matrices acting on
some finite-dimensional linear space, there exists always a basis such that
all members of this system are diagonal in this very basis if and only if
they commute with each other. Consequently, there must exist unitary and
orthogonal transformations $U_{\rm F}$ and $O_{\rm B}$ on the representation
spaces $V_{\rm F}$ and $V_{\rm B}$, respectively, such that both $y_{\rm F}$
and $C_{\rm F}$, on the one hand, as well as $y_{\rm B}$ and $C_{\rm B}$, on
the other hand, are diagonalizable simultaneously. This property of
diagonalizability is, of course, transfered to the matrices $W_I$ and $Z_\mu$
introduced in Eq.~(\ref{fin24}); in the course of this, the transformations
$U_{\rm F}$ and $O_{\rm B}$ become explicitly
\bea
U_{\rm F}&=&\dirsum{I}\,T_{\rm F}^I\times\op{1}_{d_I}\ ,\nonumber\\[1ex]
O_{\rm B}&=&\dirsum{\mu}\,T_{\rm B}^\mu\times\op{1}_{d_\mu}\ ,
\label{fin26}
\eea
where each of the transformations $T_{\rm F}^I$ serves to diagonalize a
certain isotypical block $W_I$ in $y_{\rm F}$ while each of the
transformations $T_{\rm B}^\mu$ serves to diagonalize a certain isotypical
block $Z_\mu$ in $y_{\rm B}$. By applying this diagonalization procedure to
the operators (\ref{fin5}) and (\ref{fin22}), we thus obtain, in the
fermionic sector,
\bea
{(y_{\rm F})^i}_j&=&{\d^i}_j\,y_{\rm F}^j
=\sum_{\b=1}^{d_{\rm B}}{\left(Y^{\dg\b}\,Y_{\b}\right)^i}_j\ ,\nonumber\\[1ex]
{(C_{\rm F})^i}_j&=&{\d^i}_j\,C_{\rm F}^j\ ,
\label{fin27f}
\eea
and, in the bosonic sector,
\bea
2\,{(y_{\rm B})^\a}_\b&=&2\,{\d^\a}_\b\,y_{\rm B}^\b
=\tr_{\rm F}\left(Y^{\dg\a}\,Y_{\b}+Y^{\dg\b}\,Y_{\a}\right)\ ,\nonumber\\[1ex]
{(C_{\rm B})^\a}_\b&=&{\d^\a}_\b\,C_{\rm B}^\b\ .
\label{fin27b}
\eea

As already mentioned, the YFC is invariant under all our unitary and
orthogonal transformations. Furthermore, the relations $U_{\rm F}\,R_{\rm
F}\,U_{\rm F}^\dg=R_{\rm F}$ and $O_{\rm B}\,R_{\rm B}\,\tp{O_{\rm B}}=R_{\rm
B}$ guarantee that upon application of the above diagonalization procedure
$R_{\rm F}$ and $R_{\rm B}$ remain blockdiagonal with respect to each
irreducible representation $R_{\rm F}^I\subset R_{\rm F}$ as well as with
respect to each orthogonal block $R_{\rm B}^\mu\subset R_{\rm B}$. With the
above sets of decompositions (\ref{fin27f}) and (\ref{fin27b}), the YFC
(\ref{fin10}) assumes what is usually called its standard
form:\footnote{\normalsize\ This result coincides with the well-known
standard form of the YFC but, in contrast to Ref.~\cite{kranner91}, the
simultaneous diagonal form of the operators $y_{\rm F}$, $C_{\rm F}$, $y_{\rm
B}$, and $C_{\rm B}$ was derived here without making use of the YFC.}
\be
4\sum_{\b=1}^{d_{\rm B}}\left(Y_\b\,Y^{\dg\a}\,Y_\b\right)_{ij}
+Y_{\a ij}\left(2\,y_{\rm B}^\a+y_{\rm F}^i+y_{\rm F}^j-6\,g^2\,C_{\rm F}^i
-6\,g^2\,C_{\rm F}^j\right)=0\ .
\label{fin28}
\ee
We conclude that this standard form of the YFC is quite naturally related to
a basis where both $R_{\rm F}$ and $R_{\rm B}$ are blockdiagonal.

\section{$F^2=1$ Theories}\label{fin28a}

In Ref.~\cite{kranner91} a certain---upon application of the two-loop
gauge-coupling finiteness condition, Eq.~(\ref{fin9}), purely
group-theoretic---quantity called $F$, defined by
\be
F^2:=\frac{E(Y)}{36\,g^4\,d_{\rm g}\,Q_{\rm F}}
=\frac{Q_{\rm F}+Q_{\rm B}+c_{\rm g}\,(S_{\rm F}-2\,c_{\rm g})}
{3\,Q_{\rm F}}\ ,
\label{fin29}
\ee
has been introduced. Remarkably, all theories which satisfy the central part
of finiteness conditions as represented by Eqs.~(\ref{fin8}), (\ref{fin9}),
and (\ref{fin10}) also satisfy the inequality $F\leq 1$. In particular, the
extremum $F=1$ seems to play a decisive r\^ole in the analysis of these
finiteness conditions \cite{kranner91}:
\begin{itemize}
\item If and only if this quantity $F$ is restricted to the value $F=1$, the
(cubic) YFC (\ref{fin10}) is equivalent to the (quadratic) ``$F=1$ system''
\bea
\sum_{\b =1}^{d_{\rm B}}\left(Y_{\b ij}\,Y_{\b kl}+Y_{\b ik}\,Y_{\b jl}
+Y_{\b il}\,Y_{\b jk}\right)&=&0\quad\forall\ i,j,k,l\ ,\nonumber\\[1ex]
\sum_{\a =1}^{d_{\rm B}}Y^{\dg\a}\,Y_\a&=&6\,g^2\,C_{\rm F}\ ,\nonumber\\[1ex]
\tr_{\rm F}\left(Y^{\dg\a}\,Y_\b\right)
&=&\tr_{\rm F}\left(Y^{\dg\b}\,Y_\a\right)\quad\forall\ \a,\b\ .
\label{fin30}
\eea
\item All $N=1$ supersymmetric finite theories have $F=1$ and are thus
solutions to the system (\ref{fin30}).
\item The incorporation of all supersymmetric finite theories, numerical
checks, and the fact that, in contrast to the YFC (\ref{fin10}) which is
cubic in $Y$, the system (\ref{fin30}) is only quadratic in $Y$ led to the
conjecture that all finite theories satisfy $F=1$ and belong to the solutions
of the system (\ref{fin30}).
\end{itemize}

By exploiting the highly symmetric structure of the $F=1$ system but ignoring
the requirements imposed by gauge invariance, a class of explicit solutions
of this system has been found; all members of this class are characterized by
the fact that $R_{\rm F}$ is the direct sum of merely one type of irreducible
representation while the involved Yukawa couplings are isomorphic to
generators of (a representation of) a Clifford algebra with identity element
\cite{kranner91}. In this class of theories, the ratio of the ``bosonic''
dimension $d_{\rm B}$ and the ``fermionic'' dimension $d_{\rm F}$ is
restricted to values like $d_{\rm B}/d_{\rm F}=\frac{3}{2}$, as is realized,
for instance, in all $N=4$ supersymmetric theories (which, in fact, also
exhibit a certain Clifford-like structure in their Yukawa couplings
\cite{boehm87}).

However, the construction of all these particular Clifford-like solutions of
the YFC (\ref{fin10}) takes into account neither the one-loop gauge-coupling
finiteness condition (\ref{fin8}) nor the restrictions (\ref{fin12}) on the
Yukawa couplings due to gauge invariance of the theory. The present analysis
aims at the systematic investigation of the consequences of a Clifford-like
structure of the Yukawa couplings $Y$ for finiteness of general gauge
theories.

\section{Reducibility of the Yukawa Finiteness Condition}\label{sec:reduc}

Let us now focus our attention to the standard form (\ref{fin28}) of the YFC,
obtained under the constraints (\ref{fin27f}) and (\ref{fin27b}). We notice
that $y_{\rm B}^\a$ is nothing else but the Hilbert--Schmidt norm of the
matrix $Y_\a=\tp{Y_\a}$ and that $y_{\rm F}^i$ may be interpreted as the
Hilbert--Schmidt norm of some $d_{\rm F}\times d_{\rm B}$ matrix, say $A^i$,
formed by $Y_\a$. Thus, $y_{\rm B}^\a=0$ implies that $Y_\a$ is the null
matrix, and $y_{\rm F}^i=0$ implies that $A^i$ is the null matrix.
Consequently, for a vanishing $y_{\rm B}^\a$, there cannot arise any
contributions to the YFC from $Y_{\a ij}$ for all $i,j\in\{1,\dots,d_{\rm
F}\}$, and, for a vanishing $y_{\rm F}^i$, there cannot arise any
contributions to the YFC from $Y_{\a ij}$ for all $\a\in\{1,\dots,d_{\rm
B}\}$ and for all $j\in\{1,\dots,d_{\rm F}\}$. For precisely this reason, we
find it very convenient to re-order the two bases $B_{\rm F}$ and $B_{\rm B}$
of the representation spaces $V_{\rm F}$ and $V_{\rm B}$, respectively,
according to the following
\bd
\beax
i\in\{1,\dots,n\}&\lra&y_{\rm F}^i\neq 0\ ,\\[1ex]
i\in\{n+1,\dots,d_{\rm F}\}&\lra&y_{\rm F}^i=0\ ,\\[1ex]
\a\in\{1,\dots,m\}&\lra&y_{\rm B}^\a\neq 0\ ,\\[1ex]
\a\in\{m+1,\dots,d_{\rm B}\}&\lra&y_{\rm B}^\a=0\ .
\eeax
\label{red1}
\ed
Due to Schur's lemma, this rearrangement of indices does not affect the block
structure of $R_{\rm F}$ or $R_{\rm B}$ because, as expressed by
Eq.~(\ref{fin24}), both $y_{\rm F}$ and $y_{\rm B}$ are proportional to unity
on each of the irreducible blocks given in Eq.~(\ref{fin23}) since, according
to Eq.~(\ref{fin25}), they form invariant operators acting on $V_{\rm F}$ and
$V_{\rm B}$, respectively.\footnote{\normalsize\ $y_{\rm B}$ is proportional
to unity on whole orthogonal blocks $R_{\rm B}^\mu\simeq R_{\rm B}^A\oplus
(R_{\rm B}^A)^{\rm c}$. Thus, the norm of $Y_\a$ on $R_{\rm B}^A$ equals its
norm on $(R_{\rm B}^A)^{\rm c}$!} This rearrangement procedure reduces the
YFC (\ref{fin10}) to a new system of equations. For $\a\in\{m+1,\dots,d_{\rm
B}\}$, the couplings $Y_\a$ do not contribute to this new system. For
$\a\in\{1,\dots,m\}$, all $Y_\a$ are of the form $Y_\a=E\,Y_\a$, with
projectors $E$ onto the subspace of $V_{\rm F}$ with non-vanishing $y_{\rm
F}^i$. Therefore, the YFC will involve only quantities with indices which
correspond to $y_{\rm F}^i\neq 0$ and $y_{\rm B}^\a\neq 0$:
\bea
&&4\sum_{\b=1}^{m}\left(Y_\b\,Y^{\dg\a}\,Y_\b\right)_{ij}+Y_{\a ij}
\left(2\,y_{\rm B}^\a+y_{\rm F}^i+y_{\rm F}^j
-6\,g^2\,C_{\rm F}^i-6\,g^2\,C_{\rm F}^j\right)=0\ ,\nonumber\\[1ex]
&&{\d^i}_j\,y_{\rm F}^j=\sum_{\b =1}^{m}{\left(Y^{\dg\b}\,Y_\b \right)^i}_j\ ,
\nonumber\\[1ex]
&&2\,{\d^\a}_\b\,y_{\rm B}^\b=
\tr_{\rm F}\left(Y^{\dg\a}\,Y_\b+Y^{\dg\b}\,Y_\a \right)\ ,
\label{red3}
\eea
for all $i,j\in\{1,\dots,n\}$ and for all $\a,\b\in\{1,\dots,m\}$. This new
system of equations is, of course, of the same structure as the one derived
in Sec.~\ref{sec:yfc}; however, here the Yukawa couplings $Y_{\a ij}$
contribute only for $\a\in\{1,\dots,m\}$ and $i,j\in\{1,\dots,n\}$.
Similarly, for the bounds \cite{kranner91} on the quantity $E(Y)$ of
Eq.~(\ref{eq:e(y)}), the same restricted range of fermionic indices as for
the system (\ref{red3}) is relevant:
\be
\sum_{i=1}^{n}(y_{\rm F}^i)^2\leq
E(Y)=6\,g^2\sum_{i=1}^nC_{\rm F}^i\,y_{\rm F}^i\leq
36\,g^4\sum_{i=1}^n(C_{\rm F}^i)^2\ .
\label{red4}
\ee
Hence, we encounter some fundamental difference between, on the one hand, the
full particle content of the Lagrangian (\ref{fin1}), which enters in all
group-theoretic quantities like $S_{\rm F}$, $S_{\rm B}$, $Q_{\rm F}$, or
$Q_{\rm B}$, and, on the other hand, the subset of only those particles which
also have a non-vanishing Yukawa coupling.

Just as the constraint $F=1$ can be expressed by requiring $y_{\rm
F}^i=6\,g^2\,C_{\rm F}^i$ for all $i\in\{1,\dots,n=d_{\rm F}\}$, we may set
$y_{\rm F}^i=6\,g^2\,C_{\rm F}^i$ for all $i\in\{1,\dots,n<d_{\rm F}\}$ and
get a system of the form (\ref{fin30}) with $F<1$. For this, the existence of
potentially finite theories solving Eqs.~(\ref{fin8}) and (\ref{fin9}) may be
shown numerically.

In order to construct invariant tensors for the Yukawa couplings, we
decompose both the bosonic index $\a$ and the fermionic index $i$ into pairs
of indices, say $\a=(A,\a_A)$ and $i=(I,i_I)$, where the indices $A$ and $I$
serve to distinguish irreducible representations $R_{\rm B}^A\subset R_{\rm
B}$ and $R_{\rm F}^I\subset R_{\rm F}$, respectively, while the indices
$\a_A=1,\dots,d_A$ and $i_I=1,\dots,d_I$ label the components of $R_{\rm
B}^A$ and $R_{\rm F}^I$, respectively. Let $R_{\rm B}^A\subset R_{\rm B}$,
$R_{\rm F}^I\subset R_{\rm F}$, and $R_{\rm F}^J\subset R_{\rm F}$ be three
irreducible representations of $G$. If and only if their product $R_{\rm
B}^A\otimes R_{\rm F}^I\otimes R_{\rm F}^J$ contains the trivial
representation, $\op{1}$, $N(A,I,J)$ times, there exist $N(A,I,J)$ invariant
tensors $(\Lambda^{(k)})_{\a_A i_I j_J}$. In terms of the latter, the
expansion of $Y$, with coefficients $p^{(k)}_{AIJ}\in\Bbb C$, reads
\be
Y_{\a ij}=Y_{(A,\a_A)(I,i_I)(J,j_J)}=\sum_{k=1}^{N(A,I,J)}p^{(k)}_{AIJ}
\left(\Lambda^{(k)}\right)_{\a_Ai_Ij_J}\ .
\label{red7}
\ee
We realize the naturalness of $n<d_{\rm F}$ and $m<d_{\rm B}$ in the YFC
(\ref{red3}): not all combinations of irreducible representations contained
in $R_{\rm F}$ and $R_{\rm B}$ allow to build invariant
tensors;\footnote{\normalsize\ For more details on the relation of the
expansion (\ref{red7}) and the real form of $R_{\rm B}$, see
Appendix~\ref{a8}.} every $R_{\rm F}^I$ without partners to form invariants
reduces $n$ by $d_I$, every $R_{\rm B}^A$ without partners to form invariants
reduces $m$ by $d_A$.

Now, let $M_1=\{(R^{\mu_1},R^{I_1},R^{J_1})\}$ and
$M_2=\{(R^{\mu_2},R^{I_2},R^{J_2})\}$ be two sets of combinations of real
bosonic blocks $R^{\mu_1},R^{\mu_2}\subset R_{\rm B}$ and irreducible
fermionic representations $R^{I_1},R^{I_2},R^{J_1},R^{J_2}\subset R_{\rm F}$
in the Yukawa couplings $Y_{(\mu,\a_\mu)(I,i_I)(J,j_J)}$. We define
any two sets $M_1$ and $M_2$ to be disjoint if and only if
$\{R^{\mu_1}\}\cap\{R^{\mu_2}\} = \{R^{I_1}\}\cap\{R^{I_2}\} =
\{R^{J_1}\}\cap\{R^{J_2}\} = \emptyset$.
\bd
Let $M=\{(R^\mu,R^I,R^J)\mid R^\mu\subset R_{\rm B},\ R^I,R^J\subset R_{\rm
F}\}$ be the set of all combinations of real bosonic blocks and irreducible
fermionic representations in the YFC (\ref{red3}). If $M$ is the union of
pairwise disjoint non-empty subsets $M_k$, $k=1,2,\dots,$ we call the YFC\/
{\em reducible} else\/ {\em irreducible}.\footnote{\normalsize\ Note that,
for every index of $Y_{(\mu,\a_\mu)(I,i_I)(J,j_J)}$, the splitting takes
place between the irreducible representations in $R_{\rm F}$ and real blocks
in $R_{\rm B}$. This is the finest conceivable splitting of the YFC since any
finer one would decompose $\Lambda^{(k)}$, in contradiction to
$\Lambda^{(k)}$ being a fundamental invariant tensor.}
\label{red8}
\ed

\section{Clifford Algebra Representations for Irreducible Yukawa
Finiteness Conditions}\label{sec:carfi}

For the sake of conceptual simplicity, we would like to begin the present
investigations of finiteness with the special case of an irreducible YFC. The
by far more delicate case of a reducible YFC as well as a more rigorous
treatment of the notion of reducibility of systems will be covered in
Refs.~\cite{lucha96-1,lucha96-2}.

Generalizing the ansatz which entails solutions of the YFC equivalent to
representations of some Clifford algebra \cite{kranner91}, we start with
\bd
Let the ranges of indices $n$ and $m$ be as specified in Def.~\ref{red1}.
Let the YFC be\/ {\em irreducible} in the sense of Def.~\ref{red8}. We assume
the invariant diagonalizable operator $x$ defined by Eq.~(\ref{fin14}) to be
of the form
\beax
x&=&u\otimes v\ ,\nonumber\\[1ex]
{x^{i\a}}_{j\b}&=&{u^\a}_\b\,{v^i}_j\ ,
\eeax
where $v$ and $u$ act on $V_{\rm F}$ and $V_{\rm B}$, respectively.
\label{red9}
\ed
Recalling $\tr_{\rm B}x=y_{\rm F}$ and $\tr_{\rm F}x=y_{\rm B}$ as well as
the outcome (\ref{fin27f}) and (\ref{fin27b}) of diagonalization entails, for
all $i,j\in\{1,\dots,n\}$ and for all $\a,\b\in\{1,\dots,m\}$,
\bea
&&{(\tr_{\rm B}x)^i}_j={(y_{\rm F})^i}_j={\d^i}_j\,y_{\rm F}^j=
{v^i}_j\,\tr(u)={\d^i}_j\,v^j\,\tr(u)\ ,\nonumber\\[1ex]
&&{(\tr_{\rm F}x)^\a}_\b={(y_{\rm B})^\a}_\b={\d^\a}_\b\,y_{\rm B}^\b=
{u^\a}_\b\,\tr(v)={\d^\a}_\b\,u^\b\,\tr(v)\ .
\label{cliff1}
\eea
Let us rewrite the quantities $u$ and $v$ as well as their traces in polar
decomposition:
\beax
\tr(u)&=&|\tr(u)|\exp(i\,\eta)\ ,\\[1ex]
\tr(v)&=&|\tr(v)|\exp(i\,\varphi)\ ,\\[1ex]
u^\a&=&|u^\a|\exp(i\,\eta^\a)\quad\forall\ \a\in\{1,\dots,m\}\ ,\\[1ex]
v^i&=&|v^i|\exp(i\,\varphi^i)\quad\forall\ i\in\{1,\dots,n\}\ .
\eeax
Substitution of these polar decompositions into the relations $y_{\rm
F}^i=v^i\,\tr(u)\in\Bbb R$ for all $i\in\{1,\dots,n\}$ and $y_{\rm
B}^\a=u^\a\,\tr(v)\in\Bbb R$ for all $\a\in\{1,\dots,m\}$ resulting from
Eqs.~(\ref{cliff1}) yields
\beax
\varphi^i&\equiv&-\eta\quad\forall\ i\in\{1,\dots,n\}\ ,\\[1ex]
\eta^\a&\equiv&-\varphi\quad\forall\ \a\in\{1,\dots,m\}\ .
\eeax
Moreover, because all $y_{\rm F}^i$ and all $y_{\rm B}^\a$ are real, i.e.,
$y_{\rm F}^i\in\Bbb R$ and $y_{\rm B}^\a\in\Bbb R$, we also have
\beax
\sum_{i=1}^ny_{\rm F}^i=\tr(v)\,\tr(u)\in\Bbb R\ ,\\[1ex]
\sum_{\a=1}^my_{\rm B}^\a=\tr(u)\,\tr(v)\in\Bbb R\ ,
\eeax
which, in turn, implies $\varphi\equiv -\eta$. Therefore, we end up with
\bea
u^\a&=&|u^\a|\exp(-i\,\varphi)\ ,\nonumber\\[1ex]
v^i&=&|v^i|\exp(i\,\varphi)\ ,\nonumber\\[1ex]
{x^{i\a}}_{j\b}&=&{\d^\a}_\b\,{\d^i}_j\,u^\b\,v^j=:
{\d^\a}_\b\,{\d^i}_j\,x^{j\b}\ ,
\label{cliff4}
\eea
which demonstrates that $x$ is diagonal if both $y_{\rm F}$ and $y_{\rm B}$
are diagonal. The above diagonalization of $x$ leaves the YFC unchanged; we
are thus still allowed to use the standard form of the YFC, Eq.~(\ref{red3}).
We conclude that $x=u\otimes v$ is a member of those solutions of the YFC
where $x$ is diagonalizable by some transformation of the
form\footnote{\normalsize\ These transformations $S$ correspond precisely to
the $U(d_{\rm F})\otimes O(d_{\rm B})$ symmetry of the YFC found in
Ref.~\cite{kranner91} and mentioned explicitly in Ref.~\cite{skarke94-2}.}
$S=U(n)\otimes O(m)$. (This class of solutions will be characterized in more
detail in Ref.~\cite{lucha96-2}.)

With the result (\ref{cliff4}) for $u^\a$ and $v^i$, we are able to prove
\bp
The tensorial structure of the ansatz $x=u\otimes v$ enforces a block
structure, determined by $y_{\rm F}^i$, upon $Y_\a$ for all
$\a\in\{1,\dots,m\}$:
$$
Y_{\a ij}\left(y_{\rm F}^i-y_{\rm F}^j\right)=0\quad\forall\
i,j\in\{1,\dots,n\}\ .
$$
\label{cliff5}
\ep
\prbeg We shall take repeatedly advantage of the symmetry of $Y_\a$ in its
fermionic indices, $Y_\a=\tp{Y_\a}$. We thus have
\beax
\sum_{\b=1}^m\left(Y_\b\,Y^{\dg\a}\,Y_\b\right)_{ij}=
\sum_{\b=1}^m\left(Y_\b\,Y^{\dg\a}\,Y_\b\right)_{ji}\ ,\\[1ex]
\sum_{\b=1}^m\left(Y_\b\,Y^{\dg\b}\,Y_\a\right)_{ij}=
\sum_{\b=1}^m\left(Y_\a\,Y^{\dg\b}\,Y_\b\right)_{ji}\ .
\eeax
With this and the definition (\ref{fin14}) of $x$, we find
\be
2\sum_{\b=1}^m\left[\left(Y_\b\,{x^\a}_\b\right)_{ij}-
\left(Y_\b\,{x^\a}_\b\right)_{ji}\right]=
Y_{\a ij}\left(y_{\rm F}^i-y_{\rm F}^j\right)\ .
\label{cliff7}
\ee
With the help of Eq.~(\ref{cliff4}), the two sums on the left-hand side of
Eq.~(\ref{cliff7}) may be cast into the form
$$
\sum_{\b=1}^m\left(Y_\b\,{x^\a}_\b\right)_{ij}=Y_{\a ij}\,u^\a\,v^j\ ,
$$
while, with $y_{\rm F}^i=v^i\,\tr(u)$, we have
$$
Y_{\a ij}\left(y_{\rm F}^i-y_{\rm F}^j\right)=
Y_{\a ij}\,\tr(u)\left(v^i-v^j\right)\ .
$$
Taking into account that
$$
2\,u^\a+\tr(u)=(2\,|u^\a|+|\tr(u)|)\exp(-i\,\varphi)\neq 0\ ,
$$
we obtain
$$
Y_{\a ij}\left(v^i-v^j\right)=0
$$
and, therefore,
$$
Y_{\a ij}\left(y_{\rm F}^i-y_{\rm F}^j\right)=0\ .
$$
\prend
Prop.~\ref{cliff5} may be interpreted as the alignment of ($y_{\rm
F}^i=y_{\rm F}^j$)-blocks to a blockdiagonal structure for $Y_\a$. This
structure is carried over to the YFC (\ref{red3}); it can be inserted there
to give a quasi-linear YFC:\footnote{\normalsize\ In the context of finite
quantum field theories, the notion of ``quasi-linearity'' was mentioned for
the first time in Ref.~\cite{kranner91}.}
\be
Y_{\a ij}\left(8\,x^{i\a}-2\,y_{\rm F}^i+2\,y_{\rm B}^\a-6\,g^2\,C_{\rm F}^i-
6\,g^2\,C_{\rm F}^j\right)=0\ .
\label{cliff11}
\ee
Since
\be
{x^{i\a}}_{j\a}={\left(Y^{\dg\a}\,Y_\a \right)^i}_j=
\dfrac{y_{\rm F}^i\,y_{\rm B}^\a}{{\ds\sum_{i=1}^n}y_{\rm F}^i}\,{\d^i}_j
\label{cliff12}
\ee
holds, $Y_\a$ is invertible for all $\a\in\{1,\dots,m\}$.
\br
{\em Restricting $y_{\rm F}^i$ by the two requirements $y_{\rm
F}^i=6\,g^2\,C_{\rm F}^i$ and $n=d_{\rm F}$, we recover the $F=1$ theories.
In this case, the commutator in Prop.~\ref{cliff5} is carried over to $Y_{\a
ij}\,(C_{\rm F}^i-C_{\rm F}^j)=0$, and we obtain $4+d_{\rm F}=2\,m$ and
$y_{\rm F}^i=6\,g^2\,C_{\rm F}^i=y$ for all $i\in\{1,\dots,d_{\rm F}\}$, that
is, one common value for all fermionic Casimir eigenvalues.}
\label{cliff12a}
\er

In principle, it is now straightforward to solve the YFC in the form
(\ref{cliff11}) for arbitrary values of $F$. The only quantity in
Eq.~(\ref{cliff11}) which does not depend on the Yukawa couplings $Y_{\a ij}$
is the expression $6\,g^2\,C_{\rm F}$, which is also independent of $\a$.
Furthermore, because of the (highly welcome) quasi-linearity of the YFC
(\ref{cliff11}), for this set of equations to be solvable at all, the
quantities $x^{i\a}$ must be of the order $\mbox{O}(g^2)$; that is, the
components $x^{i\a}$ of $x$, viewed as functions of $6\,g^2\,C_{\rm F}^i$,
have to be quadratic in the gauge coupling constant $g$. Beyond doubt, the
{\em ansatz}\footnote{\normalsize\ This ansatz will prove to be consistent
with the general solution of the YFC for tensorial $x=u\otimes v$
\cite{lucha96-2}.} for $x^{i\a}$ which comes first to one's mind reads
\be
x^{i\a}=6\,g^2\,a\,C_{\rm F}^i+b\quad\forall\ i\in\{1,\dots,n\}\ ,
\label{cliff13}
\ee
with arbitrary constants $a,b\in\Bbb C$. After elimination of the constant
$b$, this ansatz specifies $y_{\rm F}^i$ and $y_{\rm B}^\a$ to
\bea
y_{\rm F}^i&=&\sum_{\b=1}^mx^{i\b}=
m\,a\left(6\,g^2\,C_{\rm F}^i-\frac{6\,g^2}{n}\sum_{k=1}^nC_{\rm F}^k\right)+
\frac{1}{n}\sum_{k=1}^ny_{\rm F}^k\ ,\nonumber\\[1ex]
y_{\rm B}^\a&=&\sum_{k=1}^nx^{k\a}=\frac{1}{m}\sum_{k=1}^ny_{\rm F}^k\ .
\label{cliff14}
\eea
Substitution of these expressions into the quasi-linear YFC (\ref{cliff11})
yields
\be
Y_{\a ij}\left[6\,g^2\,[2\,a\,(4-m)-1]\,C_{\rm F}^i-6\,g^2\,C_{\rm F}^j+
2\,\frac{4-m+n}{m\,n}\sum_{k=1}^ny_{\rm F}^k-12\,g^2\,a\,\frac{4-m}{n}
\sum_{k=1}^nC_{\rm F}^k\right]=0\ ,
\label{cliff15}
\ee
which, depending on the particular value of the constant $a$ in
Eq.~(\ref{cliff13}), allows for exactly three types of solutions. For $a\neq
0$, the commutator in Prop.~\ref{cliff5} entails
\be
Y_{\a ij}\,(C_{\rm F}^i-C_{\rm F}^j)=0\quad\forall\
i,j\in\{1,\dots,n\}\ ,
\label{cliff16}
\ee
whereas, in the case $a=0$, no such statement can be made. We summarize the
solutions in form of\footnote{\normalsize\ For a sketch of the proof, see
Appendix~\ref{a15}.}
\bp
In finite quantum field theories with Yukawa couplings satisfying the tensor
structure $x=u\otimes v$ of Def.~\ref{red9} and the ansatz
$x^{i\a}=6\,g^2\,a\,C_{\rm F}^i+b$ of Eq.~(\ref{cliff13}), all $y_{\rm F}^i$,
$i\in\{1,\dots,n\}$, necessarily assume one of the following values:
\begin{enumerate}
\item[A:] For $a=0$, there is only one common value for $y_{\rm F}^i$, which
involves the average $C_{\rm m}:=(C_{\rm F}^i+C_{\rm F}^j)/2$ of the Casimir
eigenvalues:
$$
y_{\rm F}^i\equiv y=6\,g^2\,\frac{m}{4-m+n}\,C_{\rm m}\quad\forall\
i\in\{1,\dots,n\}\ .
$$
\item[B:] For $a\neq(4 - m)^{-1}$, only one fermionic Casimir eigenvalue $C$
is allowed, that is, ${(C_{\rm F})^i}_j={\d^i}_j\,C$, and only one common
value for $y_{\rm F}^i$ is possible:
$$
y_{\rm F}^i\equiv y=6\,g^2\,\frac{m}{4-m+n}\,C\quad\forall\ i\in\{1,\dots,n\}
\ .
$$
\item[C:] For $a=(4 - m)^{-1}$, different values for $y_{\rm F}^i$ are
allowed:
$$
y_{\rm F}^i=6\,g^2\,\frac{m}{4-m}\left(C_{\rm F}^i-\frac{1}{4-m+n}
\sum_{k=1}^nC_{\rm F}^k\right)\ .
$$
\end{enumerate}
\label{cliff23}
\ep
\br{\em
\begin{enumerate}
\item\label{cliff24} We note explicitly that Prop.~\ref{cliff23} is necessary
and sufficient for finding solutions of the YFC which satisfy both
$x=u\otimes v$ and the ansatz (\ref{cliff13}). However, it does not suffice
to determine potentially finite theories since the two gauge-coupling
finiteness conditions (\ref{fin8}) and (\ref{fin9}) overdetermine the YFC by
restricting the particle content of such a theory. Formally, this fact
becomes manifest by comparison of the value of $E(Y)$ with the
group-theoretic quantity equivalent to $36\,g^4\,d_{\rm g}\,Q_{\rm F}\,F^2$:
\be
E(Y)=6\,g^2\,\sum_{i=1}^nC_{\rm F}^i\,y_{\rm F}^i=
36\,g^4\,d_{\rm g}\,Q_{\rm F}\,F^2\ ?
\label{cliff25}
\ee
\item\label{cliff26} For the purpose of solving the YFC (\ref{fin10}), at
least, it is neither necessary to demand $y_{\rm F}^i\equiv y$ for all
$i\in\{1,\dots,n\}$ nor necessary to restrict the spectrum of solutions to
$F=1$. This observation rather stresses the importance of incorporating into
an eventual proof of the necessity of $F=1$ in finite quantum field theories
the gauge invariance of $Y$ as well as the gauge-coupling finiteness
conditions (\ref{fin8}) and (\ref{fin9}).
\end{enumerate}}
\label{cliff27}
\er

We call a quantum field theory ``potentially finite'' if its particle content
fulfills both the finiteness condition (\ref{fin8}) and the inequalities
$0<F^2\leq 1$ for that quantity $F$ defined by Eq.~(\ref{fin29}), if the
anomaly index of its fermionic representation, $R_{\rm F}$, vanishes, if its
bosonic representation, $R_{\rm B}$, is real, $R_{\rm B}\simeq R_{\rm B}^*$,
and if, at least, one fundamental invariant tensor, required for the
decomposition (\ref{red7}) of $Y_{\a ij}$, exists.

In view of the structure of the quasi-linear YFC (\ref{cliff11}), the ansatz
(\ref{cliff13}) for $x^{i\a}$ is independent of $\a$:
\be
{x^{i\a}}_{j\b}={\d^\a}_\b\,{\d^i}_j\,x^{j\b}=:{\d^\a}_\b\,{\d^i}_j\,x^{j}\ .
\label{cliff28}
\ee
Moreover, in our analysis of the system (\ref{red3}) only nonvanishing
$y_{\rm F}^i$, i.e., $y_{\rm F}^i\neq 0$, enter. Hence, we may divide
Eq.~(\ref{fin14}) by $x^{j}$ in order to get
\be
M^{\dg\a}\,M_\b+M^{\dg\b}\,M_\a=2\,{\d^\a}_\b\,\op{1}_n\quad\forall\
\a,\b\in\{1,\dots,m\}\ .
\label{cliff29}
\ee
Mimicking a proof given in Ref.~\cite{kranner90}, we show, in
Appendix~\ref{a22}, that any set of matrices $M_\a$ satisfying these
relations is equivalent to the union of the $n\times n$ unit matrix
$\op{1}_n$ and the subset
\be
\fb_m=\{N_\a\mid\akomm{N_\a}{N_\b}=2\,\d_{\a\b}\,\op{1}_n,\
N_{\a ij}=N_{\a ji}\in{\Bbb R},\ \a=1,\dots,m-1\}
\ee
of real, symmetric, and anticommuting elements $N_\a$ of a representation of
some Clifford algebra $\fc$:
\be
\{Y_\a,\ \a=1,\dots,m\}\sim\{\op{1}_n\}\cup{\fb}_m\ .
\label{cliff30}
\ee
\br
{\em According to Remark \ref{cliff27}.\ref{cliff26}, $F=1$ is not necessary
to allow for solutions of the YFC (\ref{fin10}) which are equivalent to
representations of Clifford algebras. Moreover, considering Case C of
Prop.~\ref{cliff23}, even solutions for different $y_{\rm F}^i$ are
possible.}
\label{cliff31}
\er

At this point, the restriction to an irreducible YFC becomes important. As a
consequence of this irreducibility assumption, the fermionic dimension $n$ of
the YFC has to coincide with the dimension of the Clifford algebra
representation. We may even use (reducible) representations of different
Clifford algebras $\fc_{p_i}$ with $\mbox{rank}\,\fc_{p_i}=p_i$ if the number
$q_i$ of elements in $\fc_{p_i}$ belonging to $\fb_m$ is large enough:
$$
m-1\leq\min_i q_i\ .
$$
The rank $p_i$ of a Clifford algebra is either even, $p_i=2\,\nu_i$, or odd,
$p_i=2\,\nu_i+1$, with $\nu_i\in\Bbb N$. If $p_i=2\,\nu_i$, then $\fc_{p_i}$
is simple and its representations are isomorphic to the direct sums of
$2^{\nu_i}\times 2^{\nu_i}$ matrices \cite{riesz93}. These matrices may be
constructed by Kronecker products of Pauli matrices \cite{boerner55}.
Exactly one half of them is totally symmetric, as required for ${\fb}_m$.
However, for $p_i=2\,\nu_i$, an additional symmetric basis element of the
Clifford algebra, the product of all generators, exists, yielding
$q_i=\nu_i+1$ symmetric anticommuting elements.\footnote{\normalsize\ The
matrices $M_\a$ satisfying Eq.~(\ref{cliff29}) transform like bi-vectors
under a change of basis. Therefore, the matrices $N_\a$ are also bi-vectors.
This behaviour under basis transformations guarantees that just the symmetric
and anticommuting elements of a Clifford algebra representation are relevant
for ${\fb}_m$.} If $p_i=2\,\nu_i+1$, then $\fc_{p_i}$ is the direct sum of
two two-sided ideals and there exist again $q_i=\nu_i + 1$ symmetric
anticommuting elements \cite{boerner55}. Let $k_i$ be the multiplicity of
$2^{\nu_i}$-blocks in some representation covering ${\fb}_m$. Then, with
$\nu:={\min_i}\,\nu_i$, $n$ must satisfy the inequality
\be
n=\sum_ik_i\,2^{\nu_i}\geq 2^\nu\,\sum_ik_i\geq 2^\nu\geq 2^{m-2}\ .
\label{cliff32}
\ee
Of course, this is only a necessary condition for a set of matrices to be
equivalent to a Clifford algebra representation. For our purposes, however,
it suffices. The actual restrictivity of this inequality may be demonstrated
by applying it directly to the class of $F=1$ theories (cf.
Remark~\ref{cliff12a}), which entails\footnote{\normalsize\ For the proof,
see Appendix~\ref{a12}.}
\bp
There exist no potentially finite $F=1$ solutions of the quasi-linear YFC
(\ref{cliff11}) which simultaneously obey the inequality (\ref{cliff32}).
\label{cliff33}
\ep
This means that Clifford solutions of the kind conjectured in
Ref.~\cite{kranner91} do not exist for an irreducible YFC.
\br {\em
\begin{enumerate}
\item \label{cliff34}
Regarding the conjecture \cite{kranner91} that there might be a connection
between solutions of the YFC being isomorphic to Clifford algebra
representations (in our sense) and $N=4$ supersymmetry, Prop.~\ref{cliff33}
excludes any such connection for the case of an irreducible YFC.
\item \label{cliff35}
Very crucial for the non-existence of $F=1$ Clifford solutions of an
irreducible YFC is the drastic restriction on the fermionic dimension imposed
by the inequality (\ref{cliff32}): $d_{\rm F}=2$ or $d_{\rm F}=4$.
\end{enumerate}}
\label{cliff36}
\er

\section{Numerics}\label{sec:num}

Having formulated the problem in a way accessible to systematic
investigation, we are now going to apply Props.~\ref{cliff5} and
\ref{cliff23} and the inequality (\ref{cliff32}) to gauge theories with
simple gauge group $G$. Because of the gauge invariance (\ref{fin12}) of the
Yukawa couplings $Y$, we have to make sure that a decomposition (\ref{red7})
of $Y$ into invariant tensors indeed exists. In order to list all interesting
theories, we have developed a C package \cite{package} which provides us with
all potentially finite theories for a given simple Lie algebra $\fa$. For
every potentially finite theory, this C package involves (optionally) a
function {\tt constraint} to be specified by the user, which we adopt to
filter all theories obeying Props.~\ref{cliff5} and \ref{cliff23} as well as
Eq.~(\ref{cliff32}). We confine ourselves to theories where all irreducible
representations able to evolve invariant tensors for Yukawa couplings
(together with their respective partners, if necessary) indeed
contribute.\footnote{\normalsize\ This means, we do not delete the
contribution of irreducible representations to the YFC by hand.} The C
package \cite{package} yields bosonic multiplicities $b_A^0$ and fermionic
multiplicities $f_I^0$, each of them describing the multiplicity of a certain
type of pairwise inequivalent irreducible representations. $R_{\rm F}$ and
$R_{\rm B}$ are then completely determined by $f_I^0$ and $b_A^0$:
\beax
R_{\rm F}&=&\dirsum{I}\,f_I^0\,R^I\ ,\\[1ex]
R_{\rm B}&=&\dirsum{A}\,b_A^0\,R^A\ .
\eeax

Now, with respect to that constant $a$ in Ansatz (\ref{cliff13}),
Prop.~\ref{cliff23} suggests to analyze the cases $a\neq 0$ and $a=0$
separately:
\begin{itemize}
\item {\em Case $a\neq 0$}: For every $R^I\subset R_{\rm F}$, we have to find
those $R^J\subset R_{\rm F}$ and $R^A\subset R_{\rm B}$ which, according to
Eqs.~(\ref{fin12}) and (\ref{red7}), satisfy $R^I\otimes R^J\otimes
R^A\supset\op{1}$, and, according to Prop.~\ref{cliff5}, have
\be
C_{\rm F}^I=C_{\rm F}^J\ .
\label{num02}
\ee
Precisely the same procedure has to be applied to every $R^A\subset R_{\rm
B}$. An (admissible) irreducible non-orthogonal representation $R^A\subset
R_{\rm B}$ enforces a non-vanishing contribution of the complete real block
$R^\mu\simeq R^A\oplus (R^A)^{\rm c}$ (cf. Appendix~\ref{a8}):
\be
R^I\otimes R^J\otimes R^\mu\supset\op{1}\quad\mbox{if and only if}\quad
R^I\otimes R^J\otimes R^A\supset\op{1}\quad\mbox{or}\quad
R^I\otimes R^J\otimes (R^A)^{\rm c}\supset\op{1}\ .
\label{num02a}
\ee
Every $R^I$ and $R^\mu$ which does not satisfy both requirements
(\ref{num02}) and (\ref{num02a}) has to be deleted from $R_{\rm F}$ and
$R_{\rm B}$, respectively. This procedure yields new multiplicities $f_I$ and
$b_\mu$. The corresponding irreducible representations then fulfill
Eqs.~(\ref{num02}) and (\ref{num02a}). Furthermore, they define the subsets
\bea
R_{\rm F}^{\rm YFC}&=&\dirsum{I}\,f_I\,R^I\subset R_{\rm F}\ ,\nonumber\\[1ex]
R_{\rm B}^{\rm YFC}&=&\dirsum{\mu}\,b_\mu\,R^\mu\subset R_{\rm B}\ ,
\label{num03}
\eea
with the dimensions
\beax
n=\dim R_{\rm F}^{\rm YFC}&=&\sum_If_I\,d_I\leq d_{\rm F}\ ,\\[1ex]
m=\dim R_{\rm B}^{\rm YFC}&=&\sum_\mu b_\mu\,d_\mu\leq d_{\rm B}\ .
\eeax
The remaining $R^I$ with non-vanishing multiplicities $f_I$ have to be
searched for different Casimir eigenvalues.\footnote{\normalsize\ We are
allowed to use $n$ and $m$ as in Def.~\ref{red1} because we assume that all
irreducible representations in Eq.~(\ref{num03}) with non-vanishing
multiplicity actually contribute to the YFC.} The number of different Casimir
eigenvalues specifies whether Case B or Case C of Prop.~\ref{cliff23} is
relevant for that particular theory. Having decided which case is actually
realized, we compute $F^2_{{\rm YFC}}$, the value of $F^2$ resulting from the
YFC. With $m$ and $n$ as given above and the abbreviations
\beax
Q_{\rm F}&=&\sum_{R^I\subset R_{\rm F}}f_I^0\,S_I\,C_I\ ,\\[1ex]
Q_{\rm YFC}&=&\sum_{R^I\subset R_{\rm F}^{\rm YFC}}f_I\,S_I\,C_I\ ,\\[1ex]
S_{\rm YFC}&=&\sum_{R^I\subset R_{\rm F}^{\rm YFC}}f_I\,S_I\ ,\\[1ex]
C^0&=&\frac{1}{4-m+n}\sum_{R^I\subset R_{\rm F}^{\rm YFC}}f_I\,d_I\,C_I=
\frac{S_{\rm YFC}\,d_{\rm g}}{4-m+n}\ ,
\eeax
we get, if $C_{\rm F}^I=C$ for all $f_I\neq 0$,
$$
F^2_{{\rm YFC}}=\frac{m}{4-m+n}\,\frac{Q_{{\rm YFC}}}{Q_{\rm F}}\ ,
$$
else
$$
F^2_{{\rm YFC}}=\frac{m}{4-m}\,\frac{Q_{{\rm YFC}}-C^0\,S_{\rm YFC}}
{Q_{\rm F}}\ .
$$
The subroutine {\tt constraint} also yields the value of $F^2$ which results
from the particle content of the theory and which may be compared with the
above $F^2_{{\rm YFC}}$:
\be
F^2_{{\rm YFC}}=\frac{Q_{\rm F}+Q_{\rm B}+c_{\rm g}\,(S_{\rm F}-2\,c_{\rm g})}
{3\,Q_{\rm F}}\ ?
\label{num4}
\ee
All theories giving equality may be regarded as good candidates for finite
quantum field theories in the sense of Prop.~\ref{cliff23}. As final check,
we apply Eq.~(\ref{cliff32}) to theories passing the criterion (\ref{num4}).
\item {\em Case $a\equiv 0$}: According to Prop.~\ref{cliff23}, let $(C_{\rm
F}^I,C_{\rm F}^J)$ for $R^I,R^J\subset R_{\rm F}$ be a pair of Casimir
eigenvalues which may couple invariantly, and let $C_{\rm m}=(C_{\rm
F}^I+C_{\rm F}^J)/2$. $R^I$ and $R^J$, and every $R^\mu\subset R_{\rm B}$
with $R^I\otimes R^J\otimes R^\mu\supset\op{1}$, contribute to the YFC. If
there exist further pairs $(C_{\rm F}^K,C_{\rm F}^L)\neq(C_{\rm F}^I,C_{\rm
F}^J)$ for $R^K,R^L\subset R_{\rm F}$, with the same $C_{\rm m}$, which allow
for invariant couplings, we collect all contributing irreducible
representations in form of
\bea
R_{\rm F}^{\rm YFC}(C_{\rm m})&=&\dirsum{I}\,f_I\,R^I\ ,\nonumber\\[1ex]
R_{\rm B}^{\rm YFC}(C_{\rm m})&=&\dirsum{\mu}\,b_\mu\,R^\mu\ ,
\label{num05}
\eea
with $f_I\neq 0$ or $b_\mu\neq 0$ if and only if $R^I\subset R_{\rm F}^{\rm
YFC}$ and $R^\mu\subset R_{\rm B}^{\rm YFC}$ exist such that $R^I\otimes
R^J\otimes R^\mu\supset\op{1}$ and $C_{\rm m}=(C_{\rm F}^I+C_{\rm F}^J)/2$.
All pairs which contribute are relevant for the computation of $F^2_{{\rm
YFC}}$:
\be
F^2_{{\rm YFC}}=\frac{m}{4-m+n}\,\frac{C_{\rm m}\,S_{\rm YFC}}{Q_{\rm F}}=
\frac{Q_{\rm F}+Q_{\rm B}+c_{\rm g}\,(S_{\rm F}-2\,c_{\rm g})}{3\,Q_{\rm F}}\ ?
\label{num5}
\ee
Finally, Eq.~(\ref{cliff32}) has to be checked. This procedure has to be
applied to all values of $C_{\rm m}$ allowed by $R_{\rm F}$.
\end{itemize}
\br {\em Our analysis is based on the standard form (\ref{red3}) of the YFC,
which is naturally related to a basis where $R_{\rm F}$ and $R_{\rm B}$ are
blockdiagonal and the invariant operators $y_{\rm F}$ and $y_{\rm B}$ are
proportional to unity in all irreducible blocks $R^I\subset R_{\rm F}$ and
$R^\mu\subset R_{\rm B}$, respectively. If, for some $R^I\subset R_{\rm F}$
or $R^\mu\subset R_{\rm B}$, no invariant tensor exists then $\left.y_{\rm
F}\right|_{\op{1}_{f_I}\times R^I}$ or $\left.y_{\rm
B}\right|_{\op{1}_{b_\mu}\times R^\mu}$, respectively, vanishes; the
non-existence of invariant tensors in some type of irreducible block means
vanishing of $\left.y_{\rm F}\right|_{\op{1}_{f_I}\times R^I}$ or
$\left.y_{\rm B}\right|_{\op{1}_{b_\mu}\times R^\mu}$.}
\label{num7}
\er

Rather surprisingly, the numerical check of the constraints (\ref{num4}) and
(\ref{num5}) for all simple Lie algebras
${\fa}=(A_r,B_r,C_r,D_r,E_6,E_7,E_8,F_4, G_2)$ up to
$r=\mbox{rank}\,{\fa}\leq 8$ produces a negative result: for $x=u\otimes v$
as in Def.~\ref{red9} and the ansatz (\ref{cliff13}) for $x^{i\a}$, there
does not exist any potentially finite theory with Yukawa couplings satisfying
an irreducible YFC if all irreducible representations allowing for invariant
tensors for the Yukawa couplings really contribute.

\section{Summary, Conclusions, and Outlook}\label{sec:sco}

Motivated by recent findings in the analysis of $F^2=1$ theories
\cite{kranner91}, we discussed particular properties and solutions of the
one-loop finiteness condition for the Yukawa couplings in general
renormalizable quantum field theories. Apart from the re-derivation of the
standard form (\ref{fin28}) of the YFC on a more fundamental level, we worked
out the importance of distinguishing carefully between the full particle
content of a theory under consideration, on the one hand, and the degrees of
freedom which actually contribute to the system (\ref{red3}), on the other
hand. The standard form (\ref{fin28}) of the YFC turns out to be merely the
consequence of the bi-linearity of the YFC and its invariance under gauge
transformations. A comprehensive characterization of this standard form is
provided by blockdiagonality of $R_{\rm F}$ in each irreducible
representation and of $R_{\rm B}$ in each real block. Demanding $y_{\rm
F}^i=6\,g^2\,C_{\rm F}^i$ for all $i\in\{1,\dots,n\}$ suffices to reduce the
(troublesome) cubic YFC (\ref{red3}) to a quadratic system of the ``$F=1$
form'' (\ref{fin30}).

The crucial observation leading to our notion of ``reducibility'' of the YFC
in the sense of Def.~\ref{red8} was that, in general, $R_{\rm F}$ and $R_{\rm
B}$ may contain subsets of irreducible representations which completely
decouple from each other. Our intention is to examine the existence of
Clifford-like Yukawa couplings in finite theories, first, by considering an
irreducible YFC. For $F=1$, the situation is summarized in
\bt
Let the YFC be irreducible, and assume $x=u\otimes v$. Then there does not
exist any $F=1$ solution of the YFC obeying the following criteria:
\begin{enumerate}
\item The fermionic representation $R_{\rm F}$ has vanishing anomaly index.
\item The bosonic representation $R_{\rm B}$ is real.
\item The beta function for the gauge coupling $g$ vanishes in one-loop
approximation.
\end{enumerate}
Hence, there cannot exist any connection between $N=4$ supersymmetry and
such Clifford solutions.
\label{sum1}
\et
By means of the physically motivated ansatz (\ref{cliff13}), using our C
package \cite{package}, we were able to prove\footnote{\normalsize\ Apart
from the fact that in Theorem~\ref{sum1} not all bosonic representations
having appropriate partners in $R_{\rm F}$ are required to contribute,
Theorem~\ref{sum3} is a generalization of Theorem~\ref{sum1}.}
\bt
Let us consider a simple Lie algebra
$$
{\fa}\in
\{A_r,B_r,C_r,D_r,E_6,E_7,E_8,F_4,G_2\mid r=\mbox{rank}\,{\fa}\leq 8\}\ ,
$$
let the YFC be irreducible, and assume $x=u\otimes v$. Then there does not
exist any solution of the YFC with
$$
x^{i\a}=6\,g^2\,a\,C_{\rm F}^i+b\quad ,\ a,b\in{\Bbb C}\ ,
$$
obeying the following criteria:
\begin{enumerate}
\item The fermionic representation $R_{\rm F}$ has vanishing anomaly index.
\item The bosonic representation $R_{\rm B}$ is real.
\item The beta function for the gauge coupling $g$ vanishes in one- and
two-loop approximation.
\item Irreducible blocks $R_{\rm B}^\mu\subset R_{\rm B}$ and $R_{\rm
F}^I,R_{\rm F}^J\subset R_{\rm F}$, with multiplicities $b_\mu$, $f_I$, and
$f_J$, respectively, which allow for invariant couplings, i.e., $R_{\rm
B}^\mu\otimes R_{\rm F}^I\otimes R_{\rm F}^J\supset\op{1}$, contribute to the
YFC such that
$$
\left.y_{\rm F}\right|_{\op{1}_{f_I}\times R^I}\neq 0
$$
and
$$
\left.y_{\rm B}\right|_{\op{1}_{b_\mu}\times R^\mu}\neq 0\ .
$$
\end{enumerate}
\label{sum3}
\et
In order to complete the investigations started here, at least two directions
have to be pursued: First, all possibilities for a reducible YFC must be
analyzed in an identical manner; this ambitious goal will be approached in
forthcoming papers \cite{lucha96-1,lucha96-2}. Secondly, by relaxing the last
criterion in Theorem~\ref{sum3}, a search for Yukawa solutions with arbitrary
amount of contribution to the YFC should be performed.

\section*{Acknowledgements}

M.~M.~would like to thank H.~Urbantke for a fruitful discussion concerning
parts in the construction of invariants. We are also indebted to A.~Prets and
W.~Spitzer for a critical reading of the manuscript.

M.~M.~was supported by ``Fonds zur F\"orderung der wissenschaftlichen
Forschung in \"Osterreich,'' project 09872-PHY, by the Institute for High
Energy Physics of the Austrian Academy of Sciences, and by a grant of the
University of Vienna.

\newpage

\appendix

\section{Invariance of the Operator $x$ on $V_{\rm F}\times V_{\rm
B}$}\label{a1}

Let $Z$: $V_{\rm F}\times V_{\rm B}\rightarrow V_{\rm F}\times V_{\rm B}$ be
defined by its components according to
\be
{Z^{i\a}}_{j\b}:=\sum_k{\left(\ol{Y}\otimes Y\right)^{\a ik}}_{\b kj}
=\sum_k\ol{Y}^{\a ik}\,Y_{\b kj}\ ,
\label{a2}
\ee
where the dual tensor $\ol{Y}$ of $Y$, which acts on the product space
$V_{\rm B}^*\times V_{\rm F}^*\times V_{\rm F}^*$, has been introduced. From
this definition, we infer that $Z$ behaves under gauge transformations like
an operator on $V_{\rm F}\times V_{\rm B}$:
\be
{Z'^{i\a}}_{j\b}=\sum_{k,l,\c,\d}{\left(R_{\rm F}\right)^i}_k\,
{\left(R_{\rm B}\right)^\a}_\c\,{\left(R_{\rm F}^{\rm c}\right)_j}^l\,
{\left(R_{\rm B}^{\rm c}\right)_\b}^\d\,{Z^{k\c}}_{l\d}\ .
\label{a3}
\ee
The operator $x$, defined by Eq.~(\ref{fin14}), is equal to half the sum of
$Z$ and its Hermitean conjugate, $Z^\dg$:
$$
2\,x=Z+Z^\dg\ .
$$
Consequently, in order to prove the invariance of $x$ under $R_{\rm F}\otimes
R_{\rm B}$, it is sufficient to show this for $Z$.

Obviously, the gauge-transformed $Z$, $Z'$, must be related to $Y'=(R_{\rm
B}^{\rm c}\otimes R_{\rm F}^{\rm c}\otimes R_{\rm F}^{\rm c})\,Y$ according
to
$$
{Z'^{i\a}}_{j\b}=\sum_k\ol{Y}'^{\a ik}\,Y'_{\b kj}\ .
$$
However, recalling the gauge invariance of $Y$ as expressed by
Eq.~(\ref{fin12}), $Y'=Y$, we get invariance of the operator $Z$ too:
$$
{Z'^{i\a}}_{j\b}={Z^{i\a}}_{j\b}\ .
$$
With Eq.~(\ref{a3}), this observation may be rephrased in the form
$$
\sum_{k,\c}{\left(R_{\rm F}\right)^i}_k\,{\left(R_{\rm B}\right)^\a}_\c\,
{Z^{k\c}}_{j\b}=\sum_{k,\c}{Z^{i\a}}_{k\c}\,{\left(R_{\rm F}\right)^k}_j\,
{\left(R_{\rm B}\right)^\c}_\b
$$
or by the commutator
$$
\komm{R_{\rm F}\otimes R_{\rm B}}{Z}=0\ ,
$$
which makes the invariance of $Z$ with respect to $R_{\rm F}\otimes R_{\rm
B}$ (on the product space $V_{\rm F}\times V_{\rm B}$) manifest.

\section{Decomposition of $Y$ into Fundamental Invariant Tensors}\label{a8}

We owe to the reader a discussion of the precise relation of Eq.~(\ref{red7})
to the bases $B_{\rm F}$ and $B_{\rm B}$ in which $R_{\rm F}$ and $R_{\rm B}$
are of the blockdiagonal form (\ref{fin23}). Recall that Eq.~(\ref{fin23})
corresponds to a decomposition according to invariant subspaces $V_I$ and
$V_\mu$ of $V_{\rm F}$ and $V_{\rm B}$, respectively, with multiplicities
$f_I$ and $b_\mu$ \cite{boerner55,urbantke76}. Performing the unitary
transformation induced by
$$
U=\frac{1}{\sqrt{2}}\left(\begin{array}{cc}\op{1}&i\,\op{1}\\[1ex]
i\,\op{1}&\op{1}\end{array}\right)
$$
in each unitary reducible orthogonal block $R^\mu\subset R_{\rm B}$ of
Eq.~(\ref{fin23}),
\be
U^\dg\,R^\mu\,U=R^A\oplus(R^A)^{\rm c}\ ,
\label{a9a}
\ee
we may express $R_{\rm B}$ as the direct sum over irreducible representations
$R^A$ with multiplicities $b_A$. We introduce a fermionic basis $\cB_{\rm F}$
and a bosonic basis $\cB_{\rm B}$ by
\beax
\cB_{\rm F}&=&\{e_I\otimes e_{i_I}\mid I=1,\dots,\sum_{I'}f_{I'};\
i_I=1,\dots,d_I\}\ ,\\[1ex]
\cB_{\rm B}&=&\{f_A\otimes f_{\a_A}\mid A=1,\dots,\sum_{A'}b_{A'};\
\a_A=1,\dots,d_A\}\ .
\eeax
Let $R_{\rm B}^A\subset R_{\rm B}$, $R_{\rm F}^I\subset R_{\rm F}$, and
$R_{\rm F}^J\subset R_{\rm F}$ be any three irreducible representations. For
these irreducible representations, $N(A,I,J)$ invariant tensors
$\Lambda^{(k)}$ exist if and only if $R_{\rm B}^A\otimes R_{\rm F}^I\otimes
R_{\rm F}^J\supset\op{1}_{N(A,I,J)}\otimes\op{1}$. Then the expansion of the
Yukawa couplings $Y$ in the tensor basis $\cB=\{f^A\otimes e^I\otimes
e^J\otimes f^{\a_A}\otimes e^{i_I}\otimes e^{j_J}\}$, with coefficients
$p^{(k)}_{AIJ}\in\Bbb C$, assumes the form (\ref{red7}):
$$
Y_{\a ij}=Y_{(A,\a_A)(I,i_I)(J,j_J)}=\sum_{k=1}^{N(A,I,J)}p^{(k)}_{AIJ}
\left(\Lambda^{(k)}\right)_{\a_Ai_Ij_J}\ .
$$
Applying to this expansion the transformation inverse to Eq.~(\ref{a9a})
gives
$$
Y_{\a ij}=Y_{(\mu,\a_\mu)(I,i_I)(J,j_J)}=\sum_{k=1}^{N(\mu,I,J)}
p^{(k)}_{\mu IJ}\left(\Lambda^{(k)}\right)_{\a_\mu i_Ij_J}\ ,
$$
where $N(\mu,I,J)=N(A,I,J)+N(A^{\rm c},I,J)$. Hence, invariant contributions
to a fixed real block $R^\mu$ arise from $R^A$ and $(R^A)^{\rm c}$; the
splitting of the YFC into subsystems is between complete real blocks.

\section{The Three Types of Solutions of the Yukawa Finiteness
Condition}\label{a15}

We solve the quasi-linear YFC (\ref{cliff15}) for the ansatz (\ref{cliff13})
by distinguishing between the following three cases:
\begin{itemize}
\item{\em Case A:} $a=0$. From the ansatz (\ref{cliff13}), we immediately
conclude that $y_{\rm F}^i=y$ for all $i\in\{1,\dots,n\}$. With this,
Eq.~(\ref{cliff15}) yields, for all $i\in\{1,\dots,n\}$ and for all
$\a\in\{1,\dots,m\}$,
$$
m\,x^{i\a}=y_{\rm F}^i=y=6\,g^2\,\frac{m}{4-m+n}\,C_{\rm m}\ ,
$$
with $C_{\rm m}:=(C_{\rm F}^i+C_{\rm F}^j)/2$.
\item{\em Case B:} $a\neq(4-m)^{-1}$. We take advantage of the commutator
(\ref{cliff16}) in order to simplify Eq.~(\ref{cliff15}) to
\be
Y_{\a ij}\left[6\,g^2\,[(4-m)\,a-1]\,C_{\rm F}^i+\frac{4-m+n}{m\,n}
\sum_{k=1}^ny_{\rm F}^k-6\,g^2\,a\,\frac{4-m}{n}\sum_{k=1}^nC_{\rm F}^k
\right]=0\ .
\label{a17}
\ee
The invertibility of $Y_{\a ij}$ expressed by Eq.~(\ref{cliff12}) allows to
sum in Eq.~(\ref{a17}) over all $i=1,\dots,n$:
\be
\sum_{k=1}^ny_{\rm F}^k=6\,g^2\,\frac{m}{4-m+n}\sum_{k=1}^nC_{\rm F}^k\ .
\label{a18}
\ee
This intermediate result may be re-inserted into Eq.~(\ref{a17}):
$$
6\,g^2\,[(4-m)\,a-1]\,C_{\rm F}^i+\frac{6\,g^2}{n}\sum_{k=1}^nC_{\rm F}^k-
6\,g^2\,a\,\frac{4-m}{n}\sum_{k=1}^nC_{\rm F}^k=0\ ,
$$
which clearly implies $C_{\rm F}^i=C$ for all $i\in\{1,\dots,n\}$, that is,
${(C_{\rm F})^i}_j={\d^i}_j\,C$. Consequently, from Eq.~(\ref{a18}), we find,
for all $i\in\{1,\dots,n\}$,
$$
y_{\rm F}^i=y=6\,g^2\,\frac{m}{4-m+n}\,C\ .
$$
\item{\em Case C:} $a=(4-m)^{-1}$. By a line of reasoning analogous to the
one applied to Case B, we get
$$
y_{\rm F}^i=6\,g^2\,\frac{m}{4-m}\left(C_{\rm F}^i-\frac{1}{4-m+n}
\sum_{k=1}^nC_{\rm F}^k\right)\ .
$$
\end{itemize}

\section{Equivalence to the Representation of a Clifford Algebra}\label{a22}

Following Ref.~\cite{kranner90}, let us briefly demonstrate that any set of
$n\times n$ matrices $Y_\a$ satisfying the relations
\be
{\left(Y^{\dg\a}\,Y_\b+Y^{\dg\b}\,Y_\a\right)^i}_j=
2\,{\d^\a}_\b\,{\d^i}_j\,x^j\quad\forall\ \a,\b\in\{1,\dots,m\}
\label{a23}
\ee
is equivalent to a representation of a Clifford algebra. First of all, the
block structure of $Y_\a$ enforced by Prop.~\ref{cliff5} can be used to split
the fermionic representation space $V_{\rm F}$ into subspaces $V_{\rm F}^i$
with $\left.y_{\rm F}\right|_{V_{\rm F}^i}=y_{\rm F}^i$. A transformation of
$Y_\a$ of the form
$$
M_\a=D\,Y_\a\,\tp{D}\ ,
$$
which preserves the symmetry of $Y_\a$ in its fermionic indices, i.e.,
$M_\a=\tp{M_\a}$, with the diagonal matrix $D$ defined by
$$
{D^i}_j:={\d^i}_j\,{\ds\frac{1}{\sqrt[4]{x^j}}}\ ,
$$
leads to
\be
{\left(M^{\dg\a}\,M_\b+M^{\dg\b}\,M_\a\right)^i}_j=2\,{\d^\a}_\b\,{\d^i}_j
\quad\forall\ \a,\b\in\{1,\dots,m\}\ .
\label{a26}
\ee
Now, each $M_\a$ may be decomposed like $M_\a=M_\a^1+i\,M_\a^2$ with real
symmetric matrices $M_\a^1$ and $M_\a^2$. For $\a=\b$, Eq.~(\ref{a26})
implies
$$
\komm{M_\a^1}{M_\a^2}=0\ .
$$
Hence, there exists an orthogonal transformation $U_0$ such that, for a fixed
$\a_0$, $M_{\a_0}$ becomes a diagonal matrix of pure phases $\zeta_j$,
$$
M'_{\a_0}=U_0\,M_{\a_0}\,\tp{U_0}=\left({\d^i}_j\exp(i\,\zeta_j)\right)\ ,
$$
while, for $\{M'_\a\}$, the analogue of Eq.~(\ref{a26}) still holds. Since
the YFC is invariant under $U(d_{\rm F})\otimes O(d_{\rm B})$
transformations, these matrices $M'_\a$ are the solutions of the YFC
transformed by $D$ and $U_0$. We may use the invariance of the YFC under
phase transformations to rotate $M'_{\a_0}$ into the $n\times n$ unit matrix:
$M''_{\a_0}=\op{1}_n$. Combining orthogonal and phase transformations,
Eq.~(\ref{a26}) becomes
$$
M''_{\b}+\ol{M''_\b}=0\quad\forall\ \b\neq\a_0:=m\ ,
$$
where $\ol{M''_\b}$ is the matrix complex conjugated to $M''_{\b}$. This, in
turn, implies
$$
M''_{\b ij}=i\,N_{\b ij}\quad\forall\ \b\in\{1,\dots,m-1\}\ ,
$$
with real and symmetric matrices $N_\b$, i.e., $N_{\b ij}=N_{\b ji}\in{\Bbb
R}$. In terms of the latter, Eq.~(\ref{a26}) reads
\be
\akomm{N_\a}{N_\b}=2\,\d_{\a\b}\,\op{1}_n\quad\forall\ \a,\b\in\{1,\dots,m-1\}
\ .
\label{a30}
\ee

\section{Clifford Solutions of the Irreducible Yukawa Finiteness Condition
for $F=1$}\label{a12}

By a straightforward inspection, we are able to preclude the existence of
Clifford-like $F=1$ solutions of an irreducible YFC. For $F=1$,
Eqs.~(\ref{cliff11}) and (\ref{cliff12}) entail
\be
4+d_{\rm F}=2\,m\leq 2\,d_{\rm B}\ ,
\label{a13}
\ee
which tells us that $d_{\rm F}$ must be even. This relation for $m$ and the
inequalities (\ref{cliff32}) conspire to restrict the possible values of
$d_{\rm F}$:
$$
d_{\rm F}\geq n\geq 2^{m-2}=2^{d_{\rm F}/2}
$$
can be fulfilled only by $d_{\rm F}=2,3,4$. Hence, we have to investigate two
possibilities: $d_{\rm F}=2$ or $d_{\rm F}=4$. The complete list of $F=1$
theories with a particle content satisfying the one-loop finiteness condition
(\ref{fin8}) for the gauge coupling, with an anomaly-free fermionic
representation $R_{\rm F}$, and with a real bosonic representation $R_{\rm
B}$, is provided by our C package \cite{package}, the subroutine {\tt
constraint} checking for $F=1$. However, irreducible representations $R^I$
with dimensions $d_I\leq d_{\rm F}\leq 4$ exist only for the Lie algebras
${\fa}=A_1,A_2,A_3,C_2$. Merely one theory in our list, for $A_1$, is
consistent with the requirement $d_{\rm F}\leq 4$. Denoting the
$d_I$-dimensional irreducible representation $R^I$ of $A_1$ by $[d_I]$, this
theory is specified by
$$
R_{\rm F}=[4]\ ,\quad R_{\rm B}=20\,[2]\oplus 7\,[3]\ ,
$$
where direct sums are implicitly understood. For these representations,
invariant tensors to construct gauge-invariant Yukawa couplings exist only
for $[3]\otimes[4]\otimes[4]$. From the decomposition (\ref{red7}) of $Y$
into invariant tensors, $m$ may take values in $\{3,6,9,\dots,21\}$, whereas
Eq.~(\ref{a13}) for $d_{\rm F}=4$ implies $m=4$.

\end{document}